\documentclass[conference,letterpaper]{IEEEtran}

\usepackage[utf8]{inputenc} 
\usepackage[T1]{fontenc}
\usepackage{url}
\usepackage{ifthen}
\usepackage{cite}
\usepackage[cmex10]{amsmath} 
\usepackage{amsmath,amsthm,amssymb}

\usepackage{svg,tikz}
\newcommand{\mysetminus}{\hbox{\tikz{\draw[line width=0.2pt,line cap=round] (1.5pt,0) -- (0,3pt);}}}
\usepackage{mathtools, nccmath}
\DeclarePairedDelimiter{\nint}\lfloor\rceil
\usepackage{booktabs}
\usepackage{svg}
\usepackage{balance}
\begin{document}
\title{Autoregressive Belief Propagation for Decoding Block Codes}

\author{%
  \IEEEauthorblockN{Eliya Nachmani and Lior Wolf}
  \IEEEauthorblockA{Tel Aviv University \& Facebook AI Research\\
                    Email: \{eliyan, wolf\}@fb.com}

}

\maketitle

\vspace{-4cm}
\begin{abstract}
We revisit recent methods that employ graph neural networks for decoding error correcting codes and employ messages that are computed in an autoregressive manner. The outgoing messages of the variable nodes are conditioned not only on the incoming messages, but also on an estimation of the SNR and on the inferred codeword and on two downstream computations: (i) an extended vector of parity check outcomes, (ii) the mismatch between the inferred codeword and the re-encoding of the information bits of this codeword. Unlike most learned methods in the field, our method violates the symmetry conditions that enable the other methods to train exclusively with the zero-word. Despite not having the luxury of training on a single word, and the inability to train on more than a small fraction of the relevant sample space, we demonstrate effective training. The new method obtains a bit error rate that outperforms the latest methods by a sizable margin. 
\end{abstract}

\section{Introduction}

The majority of learned block code decoders follow the footsteps of traditional decoders, and add learned parameters to these. This provides solid foundations to the structure of the learned decoder, and the few cases in the literature which decoders were designed de-novo, using generic neural networks, have led to weaker results. 

In this work, we propose to enrich the decoders with an autoregressive element, which is common in many other machine learning domains. When decoding is done using belief propagation (BP), it is often the case that the decoded codeword emerges gradually through the iterations. This property is used, for example, to derive loss terms that are based not only on the final output, but also on the intermediate steps. In this work, we suggest to use the intermediate decoding as input to the learned networks.

Specifically, we propose to use a hypernetwork framework, in which conditioning is performed not only based on the computed BP messages, but also based on the current level of success in decoding. We, therefore, add inputs, such as the current decoding output, the parity check of this output, and a comparison of it to an ideal version, in which a codeword is obtained from the information bits. 

For the purpose of performing the parity check, instead of using the original parity check matrix $H$, we employ, during the iterations of the method an extended version. The extended version $H'$ is obtained by  considering all pairwise combinations of the rows of $H$. This does not, of course, change the input to the decoder, nor does it change the structure of the Trellis graph that is used to design the network.

To further improve the results, we also suggest to estimate the SNR and to condition the network on an embedding of it. This allows the network to adapt to the level of noise observed for each specific input codeword.

Taken together, the new method shows an improved performance in comparison to the state of the art deep learning methods, on a diverse family of codes (BCH, LDPC, Polar). Specifically, we present an improvement of $0.5dB$ for LDPC and BCH codes, and $1.2dB$ for a Polar code. Our code is attached as supplementary.

\section{Related Work}

Deep learning have been applied to various tasks in communication~\cite{lian2018can,weinberger2020learning}, such as modulation~\cite{ramjee2020ensemble,ramjee2019fast}, equalization~\cite{caciularu2018blind, caciularu2020unsupervised}, and MIMO detection~\cite{samuel2019learning, shlezinger2020deep}.
Deep learning techniques were also applied successfully to error correcting codes, for example, encoding~\cite{jiang2019turbo}, decoding~\cite{xiao2019finite,dorner2020wgan,buchberger2020pruning,buchberger2020learned,raviv2020data,raviv2021deep,carpi2019reinforcement} and even designing new codes~\cite{kim2018deepcode} that outperform the state of the art codes for feedback channels~\cite{ben2015gaussian,ginzach2017random}. In~\cite{gruber2017deep} a fully connected neural network was employed for decoding short Polar codes up to $n=16$ bits. The results obtained are close to the maximum a posteriori (MAP) decoding, which is the optimal performance. The main drawback is that this method works well for small codes, but cannot scale to larger block codes. Another line of work for decoding Polar codes, partitions the polar encoding graph into sub-blocks, and decodes each sub-block separately~\cite{cammerer2017scaling}. 
In~\cite{kim2018communication}, an RNN decoder for convolutional and Turbo codes was introduced, which can match the performance of the classical BCJR decoder and the Viterbi decoder.

There are several methods for decoding relatively large block codes ($n\geq 100$). The well known BP decoding algorithm was unfolded into a neural network, where the variable edges were equipped with learnable parameters~\cite{nachmani2016learning}. A hardware friendly decoder of a similar nature was then introduced~\cite{lugosch2017neural}, in which the min-sum algorithm is employed. Both of these methods show an improvement over the baseline BP algorithm. Further improvements were obtained by~\cite{nachmani2019hyper}. First, each variable node was extended to a small neural network $g$ which transforms the decoder into a  graph neural network. Second, the weights of each variable nodes network $g$ are determined by a hypernetwork~\cite{ha2016hypernetworks} $f$, which provides the neural decoder with added adaptation.

In this work, we focus on decoding the following block codes -- LDPC, BCH, and Polar, where the most relevant baselines which we improve are~\cite{nachmani2017learning, nachmani2019hyper}. This improvement is obtained mostly by introducing autoregressive signals. The method is related to the Dynamic Factor Graphs (DFG)~\cite{mirowski2009dynamic}, which improves the performance of a factor graph used for time-series analysis by conditioning the current state on the previous states. We also condition the hyeprnetwork on an estimated SNR. This is related to~\cite{goldberger2010pseudo}, which conditioned the classic BP algorithm on the output of minimum mean square error (MMSE) prediction for the task of LDPC decoding.

\section{Background}

The BP methods, which were traditionally used for decoding block code, have been augmented with learned weights, and, more recently, generalized to graph neural networks (GNN)~\cite{satorras2020neural} and even to graph hyper networks~\cite{nachmani2019hyper}. Our method relies on hyper networks as well and the background below follows the naming conventions of~\cite{nachmani2019hyper}. 

A block code has $k$ information bits and $n$ output bits. The parity check matrix $H$ that defines the code is a binary matrix of size $(n-k) \times n$;   The generator matrix $G$ is a binary matrix of size $k \times n$. The standard form of $G$ has the structure of $[I_k | P]$ where, $I_k$ is the identity matrix with a dimension of $k \times k$ and $P$ is $k \times (n-k)$ binary matrix.

The vanilla belief propagation algorithm operates on the Trellis graph, which is a directed graph that we view as a layered neural network. The input layer has $n$ nodes corresponding to each input bit.  The hidden layers are of two types, namely variable layers and check layers. These layers are interleaved, such that the layer index $j$ is odd for variable layers, and even for check layers.

Each column of the matrix $H$ is associated with one bit of the codeword and with $d_v$ variable nodes in each variable layer, where $d_v$ is the sum over this column. For notational convenience, we assume that $H$ is regular, i.e, that the sum over columns ($d_v$) is fixed. Therefore, each variable layer has  $E=d_v \cdot n$ \textit{variable processing units}.

Similarly, assuming regularity also over the rows of $H$, the check layers are composed of $E=(n-k)\times d_c$ \textit{check processing units}, each associated with a parity check. 

In the experiment section, we will show that our method also works in an irregular scenario, i.e. where each variable/check node has a different degree.

The messages propagate in the Trellis graph from a variable layer to a check layer iteratively. The input to the belief propagation algorithm is the log likelihood ratio (LLR) $\ell \in \mathbb{R}^{n} $ of each bit:
\begin{equation}
\ell_v = \log\frac{\Pr\left(c_v=0 | y_v\right)}{\Pr\left(c_v=1 | y_v\right)},
\end{equation}
where $y_v$ is the received signal which is associated with the $c_v$ bit that we wish to recover. And $\ell_v$ is the corresponding log likelihood ratio.

Let $x^j$ be the messages vector in the belief propagation algorithm, of length $E$. The input layer expands the vector $\ell$ of $n$ LLR values, to the messages vector $x^j$ with $E$ elements:
\begin{equation}
\label{eq:in_layer}
    x^j = W_{in} \cdot \ell
\end{equation}
where $W_{in}$ is a binary matrix with a dimension of $E \times n$.  $W_{in}$ is constructed from the parity check matrix $H$. Each row of $W_{in}$ is associated with a row $i$ and a column $j$ of $H$ and is set as $H_i - \delta_{j}$ if $H_{ij}=1$, where $\delta_{j}$ is a sparse vector that contain $1$ only in the $j$-th location, $H_i$ is the $i$-th row of $H$, and $H_{ij}$ is a single value of this matrix.

Each iteration $j$ of the BP algorithm produces a message $x^j$. Each element $e$ in the vector $x^j$ is given by:
\begin{equation}
\label{eq:odd}
\begin{split}
x^{j}_e = x^{j}_{(c,v)} = \tanh \left(\frac{1}{2}\left(l_v + \sum_{e'\in N(v)\setminus \{(c,v)\}} x^{j-1}_{e'}\right)\right) \\ \text{if $j$ is odd}\raisetag{22pt}
\end{split}
\end{equation}
\begin{equation}
\label{eq:even}
\begin{split}
x^{j}_e = x^j_{(c,v)} = 2 \mathrm{arctanh} \left( \prod_{e'\in N(c) \setminus \{(c,v)\}}{x^{j-1}_{e'}}\right) \\ \text{if $j$ is even}
\end{split}
\end{equation}
\noindent where $N(c)=\{(c,v) | H(c,v)=1\}$ is the set of edges in which the check node $c$ is associated with the variable node $v$ in the Trellis graph. The messages propagate in $L$ variable layers and $L$ check layers until the $2L+1$ final marginalization layer. This layer outputs the marginalization:
\begin{equation}
\label{eq:ov_inter}
    u^j_v = l_v + \sum_{e'\in N(v)} x^{j-1}_{e'}
\end{equation}

\begin{equation}
\label{eq:ov}
    o^j_v = \sigma \left( u^j_v \right)
\end{equation}
where $\sigma$ is the sigmoid function. The final marginalization can be also written in matrix form by:

\begin{equation}
\label{eq:out_form}
    u^j = x^{j-1} W_{out}
\end{equation}

 where $W_{out}$ is a binary matrix of dimensionality $E \times n$. Every group of rows in $W_{out}$ is associated with column $j$ of $H$, which is denote $H_j$. The group has as many rows as the number of ones in $H_j$ and each row is sparse and contains a single 1 element in a location $i$ that matches a value of one in $H_j$ ($H_{ij}=1$).

In~\cite{nachmani2019hyper} a learnable hypernet graph neural network is introduced. First, the vanilla belief propagation is transformed into a graph neural network, by replacing each variable node with a small learnable neural network $g$. Second, the primary network (a term from the hypernetwork literature) $f$ is used to predict the  weights of the neural network $g$. These two modifications take place by replacing Eq.~\ref{eq:odd} (odd $j$, where $j$ is the iteration number) with:
\begin{equation}
\label{eq:oddreplaced}
x^{j}_e = x^{j}_{(c,v)} = g(l_v,x^{j-1}_{N(v,\mysetminus c)},\theta_g^j),
\end{equation}
\begin{equation}
\label{eq:f}
    \theta_g^j = f(|x^{j-1}|,\theta_f)\,,
\end{equation}
where $\theta_g^j$ and $\theta_f$ are the weights of the network $g$ and $f$ respectively. $x^{j}_{N(v,\mysetminus c)}$ is a vector of the elements from $x^j$ that corresponds to the indices $N(v)\setminus \{(c,v)\}$, and has a length of $d_v-1$. The hypernet scheme is shown to increase the adaptiveness of the  network, which helps overcome errors during the decoding process.

Training is performed as in~\cite{nachmani2016learning} by observing the marginalization at each odd iteration and applying cross entropy. The marginalization at the odd layer $j$ is 
\begin{equation}
\label{eq:marg}
o^{j}_v = \sigma \left( l_v + \sum_{e'\in N(v)}\bar{w}_{e'} x^{j}_{e'} \right)
\end{equation}
where $\bar{w}_{e'}$ is learnable matrix. The loss function is:
\begin{equation} 
\mathcal{L}=-\frac{1}{n}\sum_{h=1}^{L}\sum_{v=1}^{n} c_{v}\log(o^{2h+1}_{v})+(1-c_{v})\log(1-o^{2h+1}_{v}) 
\label{eq:loss_neurips19}
\end{equation}
where $c_v$ is the vector of ground truth bits.
The problem setup assumes additive white Gaussian noise (AWGN) channel with Binary Phase Shift Keying (BPSK) modulation.

\begin{figure}
    \centering
        \includegraphics[width=.4\textwidth,height=.2\textheight,keepaspectratio]{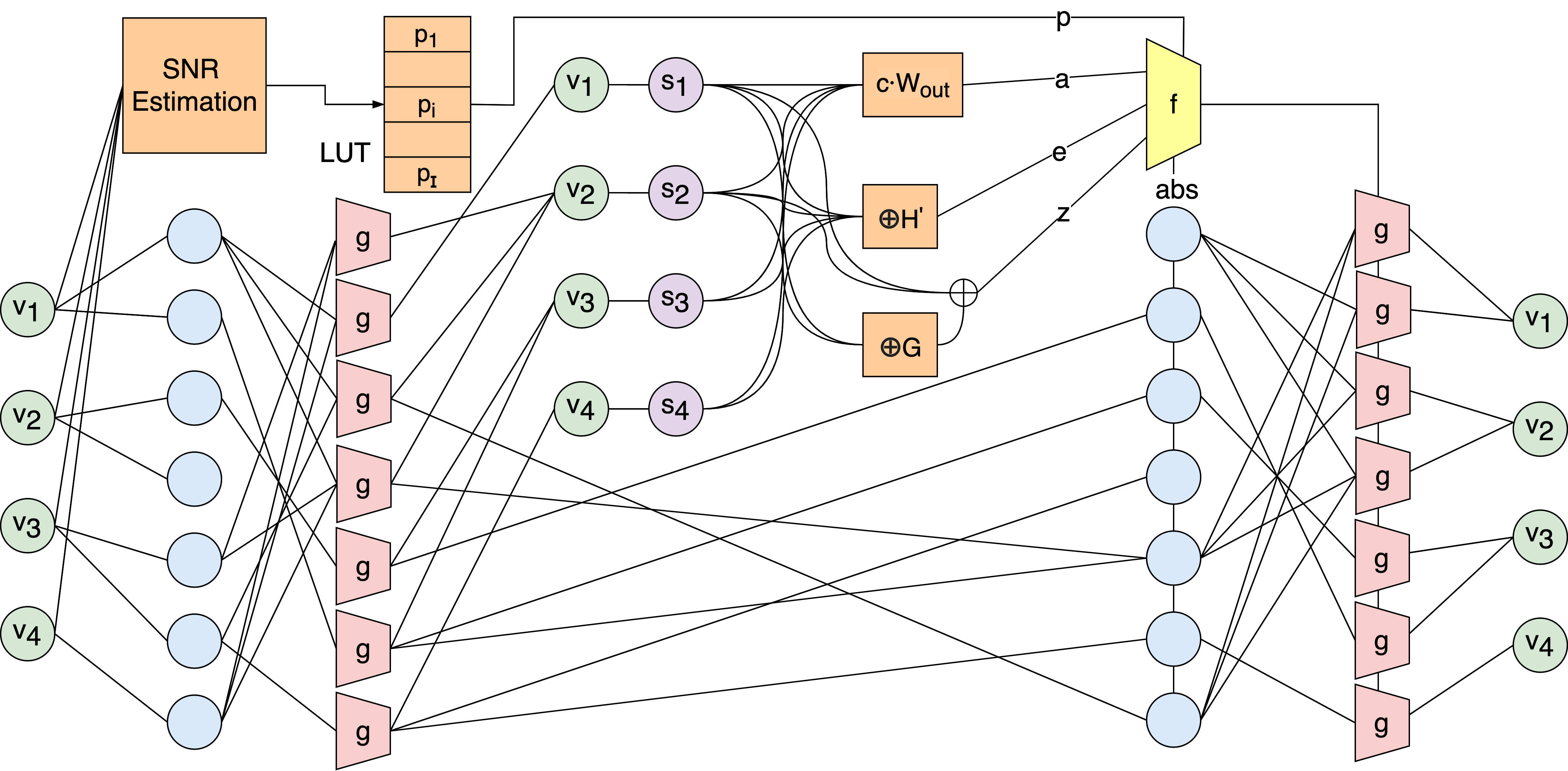}
    \caption{An overview of our method for a linear block code with $n=4$, $k=2$.}
    \label{fig:arch}
\end{figure}

\section{Method}

We suggest adding four major components to the learnable belief propagation algorithm: (i) the autoregressive signal $a^{j}$ of the current estimation of the output variable, (ii) an autoregressive vector of parity check outcomes $e^j$, based on the parity check matrix and the column combinations, (iii) an autoregressive error vector $z^j$ based on the re-encode codeword, and (iv) an embedding of Signal-To-Noise ratio $p$, see Fig.~\ref{fig:arch}. Specifically, we replace Eq.~\ref{eq:f} with the following:
\begin{equation}
\label{eq:f_new}
    \theta_g^j = f(a^{j}, e^{j}, z^{j}, p, |x^{j-1}|, \theta_f)\,.
\end{equation}
The first input $a^{j}$ is a vector of length $E$, which contains the output scaled projection of the previous $j-1$ belief propagation algorithm after Binary Phase Shift Keying (BPSK) modulation, given by:
\begin{equation}
\label{eq:ar}
    a^{j} = c_{j} \cdot (W_{out} \cdot s^{j})
\end{equation}
where $c_{j}$ is a learnable scale variable, {$W_{out}$ is defined after Eq.~\ref{eq:out_form}}, and $s^{j}$, is the hard decision vector with length of $n$:
\begin{equation}
\label{eq:slicer}
    s^j= \left\{
    \begin{matrix}
    1 & o^{j-1} > 0.5\\ 
    -1 & o^{j-1} \leqslant 0.5
    \end{matrix}\right. \,,
\end{equation}
where $o^{j}$ defined in Eq.~\ref{eq:ov} and the binarization occurs per each vector element. Note that $W_{out}$ is used to project the hard decision vector into the same dimension as $|x^{j-1}|$. 

The second input of $f$, $e^{j}$, contains information about violated parity checks. We consider an augment the parity check matrix $H'$, which extends the parity check matrix $H$, and contains all pairwise row combinations of $H$. Let $d = \binom{n-k}{2}$
, then the new parity check matrix ${H'}$ has a dimension of $d \times  n$, and is given by:
\begin{equation}
\label{eq:h_hat}
    {H'}_{\alpha\beta} =  H_\alpha \oplus H_\beta
\end{equation}
where $\oplus$ is the $XOR$ operation (multiplication modulo 2), $H'_{\alpha\beta}$ is a row of $H'$ with the double index $\alpha\beta$, and $H_\alpha$($H_\beta$) is row $\alpha$($\beta$) of $H$. In this manner, the second input to $f$, $e^{j}$ is a binary vector with a length of $d$, and is given by:
\begin{equation}
\label{eq:e_vec}
    e^{j} = {H'} \oplus s^{j} 
\end{equation}

The third inputs of $f$, $z^{j}$, is an error measurement based on the re-encoded parity check bits of the autoregressed codeword. We suggest to re-encode the parity check bits, by using the generator matrix $G$ and the information symbols of the hard decision $s^j$. Since we use $G$ in its standard form, we can regard $s^j$ as concatenation of two vectors, $s^j = [s^j_{\text{info}}, s^j_{\text{parity}}]$, of size $k$ and $n-k$, respectively, the first denoting the information bits and the second the parity check bits. 

The re-encoded codeword ${s'}^j$ of length $n$ is given by:
\begin{equation}
\label{eq:s_encode}
    {s'}^j = s^j_{\text{info}} \oplus G 
\end{equation}
In the same manner as above, we can regard ${s'}^j$ as concatenation of two vectors ${s'}^j = [{s'}^j_{\text{info}}, {s'}^j_{\text{parity}}]$.

The error measurement $z^{j}$, a vector of length $(n-k)$, is the mismatch between the autoregressed parity check bits $s^j_{\text{parity}}$ and the re-encoded ones ${s'}^j_{\text{parity}}$:
\begin{equation}
\label{eq:z_vec}
    z^{j} = s^j_{\text{parity}} \oplus {s'}^j_{\text{parity}}
\end{equation}

In order to compute the fourth component $p$, we estimate the SNR of the received signal $y$. Since the problem setup assumes an AWGN channel, the received signal is given by:
\begin{equation}
\label{eq:rec_y}
    y = \frac{2}{\sigma^2_n}\left ( {s} + \sigma_n \cdot n \right )
\end{equation}
where ${s}$ is a vector of $+1/-1$ bit symbols, $n$ is normal Gaussian noise vector and $\sigma_n$ is the noise variance which equal to $(\sqrt{2 \cdot R \cdot 10^{\frac{SNR}{10}}})^{-1}$.

An estimation $\tilde p$ of the SNR in dB given $y$ is therefore:
\begin{equation}
\label{eq:snr_est}
    \tilde{p}=10\cdot log_{10}\left ( \frac{Var^{\pm}(y)}{8 \cdot R}\right )
\end{equation}
where $Var^{\pm}(y)$ is an estimation of the variance based on a hard decision on the received signal, i.e., by thresholding the vector $y$ and partitioning it to obtain the positive elements $y^{+}$ and the negative elements $y^{-}$:
\begin{equation}
\label{eq:vary}
    Var^\pm(y) = \frac{1}{2} \cdot \left ( Var(y^{+}) + Var(y^{-}) \right )
\end{equation}

Valid SNR values are between one and $I$. The estimated value $\tilde{p}$ may be out of range of the valid SNR values and we, therefore, round and trim it, obtaining a corrected estimation $\bar p = min(max(0,\nint{\tilde p}),I)$, where $I$ is the maximal SNR value and $\nint{\cdot}$ is the $round$ operation.

$p$ is an embedding of the estimated SNR value $\bar p$. Let $i=1,2, ... , I$ be the estimated SNR values (in $dB$). Let $LUT_{snr} \in \mathbb{R}^{64 \times I}$ be the SNR look-up-table, where each column $p_i \in \mathbb{R}^{64}$ is the learned vector embedding of SNR $i$. Given an estimation $\bar p$, we simply set $p = p_{\bar p}$.

\subsection{Complexity Analysis}
We next compare the complexity of the Neural BP decoder \cite{nachmani2016learning}, the hypernetwork BP of \cite{nachmani2019hyper} and our proposed method. 
The move from Neural BP \cite{nachmani2016learning} to hypernetwork BP \cite{nachmani2019hyper} adds a complexity term of:
$O\left ( LE n_{u,g} n_{u,f}\right )$
where $L$ is the number of the iterations, $E$ is the number of processing units, and $n_{u,g}$ and $n_{u,f}$ are the number of neurons in $g$ and $f$ networks respectively. Our proposed method add on top of \cite{nachmani2019hyper} the following number of multiplication:
$O\left ( Ln\left ( E+d+n_{u,f} \right ) \right )$
where $n$ is the number of bits in the code and $d = \binom{n-k}{2}$. For example, decoding BCH(63,51) with our method adds 5\% to the total number of operations (and gains an improvement of $0.5dB$, see Sec.~\ref{sec:experiments}). This is also validated in the actual runtime, as reported in Tab.~\ref{tab:runtime}. 

However, the performance of our autoregressive method for $L=5$ iterations convincingly outperforms the method of \cite{nachmani2019hyper} for $L=50$ iterations. This $10\times$ reduction in the number of iterations is much more substantial than an increase of 5\% in the runtime per iteration.

\subsection{Symmetry conditions}
Decoding block codes with the message passing decoder that maintains the symmetry condition has the desired property that the error is independent of the transmitted codeword~\cite[Definition 4.83]{richardson2008modern}. The direct implication is that the training set can contain only noisy variations of the zero codeword. Our method does not preserve the symmetry condition and, therefore, the training set should contain random codewords. The symmetry condition for a variable node at iteration $j$ is given by:
\begin{equation}
    \label{eq:sym_var}
    \Psi\left(-l_v,-x^{j-1}_{N(v,\mysetminus c)}\right )= -\Psi\left (l_v,x^{j-1}_{N(v,\mysetminus c)}\right)
\end{equation}
where $\Psi$ is the computation in the variable node.

Assuming that the variable node calculation is given by Eq.~\eqref{eq:oddreplaced} and Eq.~\eqref{eq:f_new}, then the proposed architecture does not satisfy the variable symmetry condition. The underlying reason is that $g$ employs the anti-symmetric function $tanh$, but the inputs to this function are not sign invariant. 
Let $K=d_v-1$ and $x^{j}_{N(v,\mysetminus c)} = \left(x^{j}_{1}, \dots, x^{j}_{K} \right)$. In the proposed architecture for any odd $j\geqslant0$, $\Psi$ is given as 
\begin{equation}
\begin{aligned}
g\big(l_v,x^{j-1}_{1},\dots, x^{j-1}_{K}, \theta^{j}_g\big) =  \mathrm{tanh} \big ( W_{p}^\top  \enspace \\
... \enspace \mathrm{tanh} \big (W_{2}^\top \mathrm{tanh} \big ( W_{1}^\top \big(l_v,x^{j-1}_{1},\dots, x^{j-1}_{K}\big) \big )\big )\big )
\end{aligned}
\end{equation}
where $p$ is the number of layers and the weights $W_{1},...,W_{p}$ constitute $\theta^{j}_g=f(a^{j}, e^{j}, z^{j}, p, |x^{j-1}|,\theta_f)$. 
For real valued weights $\theta^{lhs}_g$ and $\theta^{rhs}_g$, since $\mathrm{tanh}(x)$ is an odd function, for any real value input, if  $\theta^{lhs}_g = \theta^{rhs}_g$ then
$g\left(l_v,x^{j-1}_{1},\dots, x^{j-1}_{K},\theta^{lhs}_g \right) = - g\left(-l_v,-x^{j-1}_{1},\dots, -x^{j-1}_{K},\theta^{rhs}_g \right)$. In our case, 
    $\theta^{lhs}_g = f(a^{j}, e^{j}, z^{j}, p, |x^{j-1}|,\theta_f) \neq f(-a^{j}, -e^{j}, -z^{j}, -p, |-x^{j-1}|,\theta_f) = \theta^{rhs}_g$ since $a^j=c_{j} \cdot (W_{out} \cdot s^{j}) \neq -a^j = -c_{j} \cdot (W_{out} \cdot s^{j}) = c_{j} \cdot (W_{out} \cdot (-s^{j}))$. Similar arguments hold for other terms, such as $e^j$ and $z^j$.

\subsection{Training}
Training is performed using the loss function in Eq.~\ref{eq:loss_neurips19}. Similarly to~\cite{nachmani2019hyper}, the Taylor approximation to Eq.~\ref{eq:even} is employed, in order to stabilize the training. Since the variable symmetry condition is violated, we train the model with random codewords. We use the Adam optimizer~\cite{kingma2014adam} for training, with the learning rate of $1e-4$ for all block codes. Moreover, the graph neural network simulated $L=5$ iteration of the BP algorithm, which is equal to $10$ layers. 

\begin{table}[t]
    \centering
    \caption{The negative natural logarithm of the Bit Error Rate (BER) at three SNR values. Higher is better.}
    \label{tab:ber_snr}
    \setlength\tabcolsep{1.5pt}
    \begin{tabular}{lc@{~}c@{~}cc@{~}c@{~}cc@{~}c@{~}cc@{~}c@{~}c}
    \toprule
         
        Method & \multicolumn{3}{c}{BP} &  \multicolumn{3}{c}{\cite{nachmani2017learning}} & \multicolumn{3}{c}{Hypernet} &
        \multicolumn{3}{c}{Ours}\\
        \cmidrule(lr){2-4}
        \cmidrule(lr){5-7}
        \cmidrule(lr){8-10}
        \cmidrule(lr){11-13}
         & 4 & 5 & 6 & 4 & 5 & 6 & 4 & 5 & 6 & 4 & 5 & 6\\ 
        \midrule 
        \multicolumn{13}{c}{--- after five iterations ---}\\
        Polar(64,32) & 3.52 & 4.04 & 4.48 & 4.14 & 5.32 & 6.67 & 4.25 & 5.49 & 7.02 & {\bf4.77} & {\bf6.30} & {\bf8.19} \\
        Polar(64,48) & 4.15 & 4.68 & 5.31 & 4.77 & 6.12 & 7.84 & 4.91 & 6.48 & 8.41 & {\bf5.25} & {\bf6.96} & {\bf9.00} \\
        Polar(128,64)& 3.38 & 3.80 & 4.15 & 3.73 & 4.78 & 5.87 & 3.89 & 5.18 & 6.94 & {\bf4.02} & {\bf5.48} & {\bf7.55} \\
        Polar(128,86) & 3.80 & 4.19 & 4.62 & 4.37 & 5.71 & 7.19 & 4.57 & 6.18 & 8.27 & {\bf4.81} & {\bf6.57} & {\bf9.04} \\
        Polar(128,96) & 3.99 & 4.41 & 4.78 & 4.56 & 5.98 & 7.53 & 4.73 & 6.39 & 8.57 & {\bf4.92} & {\bf6.73} & {\bf9.30}  \\
        LDPC(49,24) & 5.30 & 7.28 & 9.88 & 5.49 & 7.44 & 10.47 & 5.76 & 7.90 & 11.17 &  {\bf6.05} & {\bf8.13} & {\bf11.68} \\
        LDPC(121,60) & 4.82 & 7.21 & 10.87 & 5.12 & 7.97 & 12.22 & 5.22 & 8.29 & 13.00 & {\bf5.22} & {\bf8.31} & {\bf13.07} \\
        LDPC(121,70) & 5.88 & 8.76 & 13.04 & 6.27 & 9.44 & 13.47 & 6.39 & 9.81 & 14.04 & {\bf6.45} & {\bf10.01} & {\bf14.77} \\
        LDPC(121,80) & 6.66 & 9.82 & 13.98 & 6.97 & 10.47 & 14.86 & 6.95 & 10.68 & 15.80 & {\bf7.22} & {\bf11.03} & {\bf15.9} \\
        MacKay(96,48) & 6.84 & 9.40 & 12.57 & 7.04 & 9.67 & 12.75 & 7.19 & 10.02 & 13.16 & {\bf7.43} & {\bf10.65} & {\bf14.65} \\
        CCSDS(128,64) & 6.55 & 9.65 & 13.78 & 6.82 & 10.15 & 13.96 & 6.99 & 10.57 & 15.27 & {\bf7.25} & {\bf10.99} & {\bf16.36} \\
        BCH(31,16) & 4.63 & 5.88 & 7.60 & 4.74 & 6.25 & 8.00 & 5.05 & 6.64 & 8.80 & {\bf5.48} & {\bf7.37} & {\bf9.61} \\
        BCH(63,36) &  3.72 & 4.65 & 5.66 & 3.94 & 5.27 & 6.97 & 3.96 & 5.35 & 7.20 & {\bf4.33} & {\bf5.94} & {\bf8.21} \\
        BCH(63,45) & 4.08 & 4.96 & 6.07 & 4.37 & 5.78 & 7.67 & 4.48 & 6.07 & 8.45 & {\bf4.80} & {\bf6.43} & {\bf8.69} \\
        BCH(63,51) & 4.34 & 5.29 & 6.35 & 4.54 & 5.98 & 7.73 & 4.64 & 6.08 & 8.16 & {\bf4.95} & {\bf6.69} & {\bf9.18} \\  
        \midrule   
        \multicolumn{13}{c}{--- at convergence ---}\\
        Polar(64,32) & 4.26 & 5.38 & 6.50 & 4.22 & 5.59 & 7.30 & 4.59 & 6.10 & 7.69  & {\bf5.57} & {\bf7.43} & {\bf9.82} \\ 
        Polar(64,48) & 4.74 & 5.94 & 7.42 & 4.70 & 5.93 & 7.55 & 4.92 & 6.44 & 8.39  & {\bf5.41} & {\bf7.19} & {\bf9.30} \\ 
        Polar(128,64) & 4.10 & 5.11 & 6.15 & 4.19 & 5.79 & 7.88 & 4.52 & 6.12 & 8.25  & {\bf4.84} & {\bf6.78} & {\bf9.30} \\ 
        Polar(128,86) & 4.49 & 5.65 & 6.97 & 4.58 & 6.31 & 8.65 & 4.95 & 6.84 & 9.28  & {\bf5.39} & {\bf7.37} & {\bf10.13} \\ 
        Polar(128,96) & 4.61 & 5.79 & 7.08 & 4.63 & 6.31 & 8.54 & 4.94 & 6.76 & 9.09  & {\bf5.27} & {\bf7.44} & {\bf10.2} \\
        LDPC(49,24)& 6.23 & 8.19 & 11.72 & 6.05 & 8.34 & 11.80 & 6.23 & 8.54 & 11.95  & {\bf6.58} & {\bf9.39} & {\bf12.39} \\ 
        BCH(63,36) & 4.03 & 5.42 & 7.26 & 4.15 & 5.73 & 7.88 & 4.29 & 5.91 & 8.01 & {\bf4.57} & {\bf6.39} & {\bf8.92} \\ 
        BCH(63,45) & 4.36 & 5.55 & 7.26 & 4.49 & 6.01 & 8.20 & 4.64 & 6.27 & 8.51 & {\bf4.97} & {\bf6.90} & {\bf9.41} \\ 
        BCH(63,51) & 4.58 & 5.82 & 7.42 & 4.64 & 6.21 & 8.21 & 4.80 & 6.44 & 8.58 & {\bf5.17} & {\bf7.16} & {\bf9.53} \\ 
        \bottomrule
    \end{tabular}
\end{table} 

\section{Experiments}
\label{sec:experiments}
We trained our proposed architecture with three types of linear block codes: Low Density Parity Check (LDPC) codes~\cite{gallager1962low,shuval2010universality,soljanin2005incremental}, Polar codes~\cite{arikan2008channel,tal2017construction,liu2017polar,goela2014polar} and Bose–Chaudhuri–Hocquenghem (BCH) codes~\cite{bose1960class}. All parity check matrices and generator matrices are taken from~\cite{channelcodes}. If the generator matrices do not have a standard form, we rearrange the columns of $G$ to a standard form.

The training set contains generated random examples that are transmitted over an additive white Gaussian noise (AWGN). Each batch contains multiple sets of Signal-To-Noise (SNR) values. The same hyperparameters are used for all types of codes. We use a batch size of $120$ examples, and each batch contains $15$ examples per SNR value of $1dB,2dB,..,8dB$. The order of the Taylor approximation to Eq.~\ref{eq:even} was $q=1005$. the network $f$ has four layers with 128 neurons at each layer. The network $g$ has two layers with 16 neurons at each layer

\begin{figure}
\centering
\begin{tabular}{c@{~}c}
\includegraphics[width=.25\textwidth]{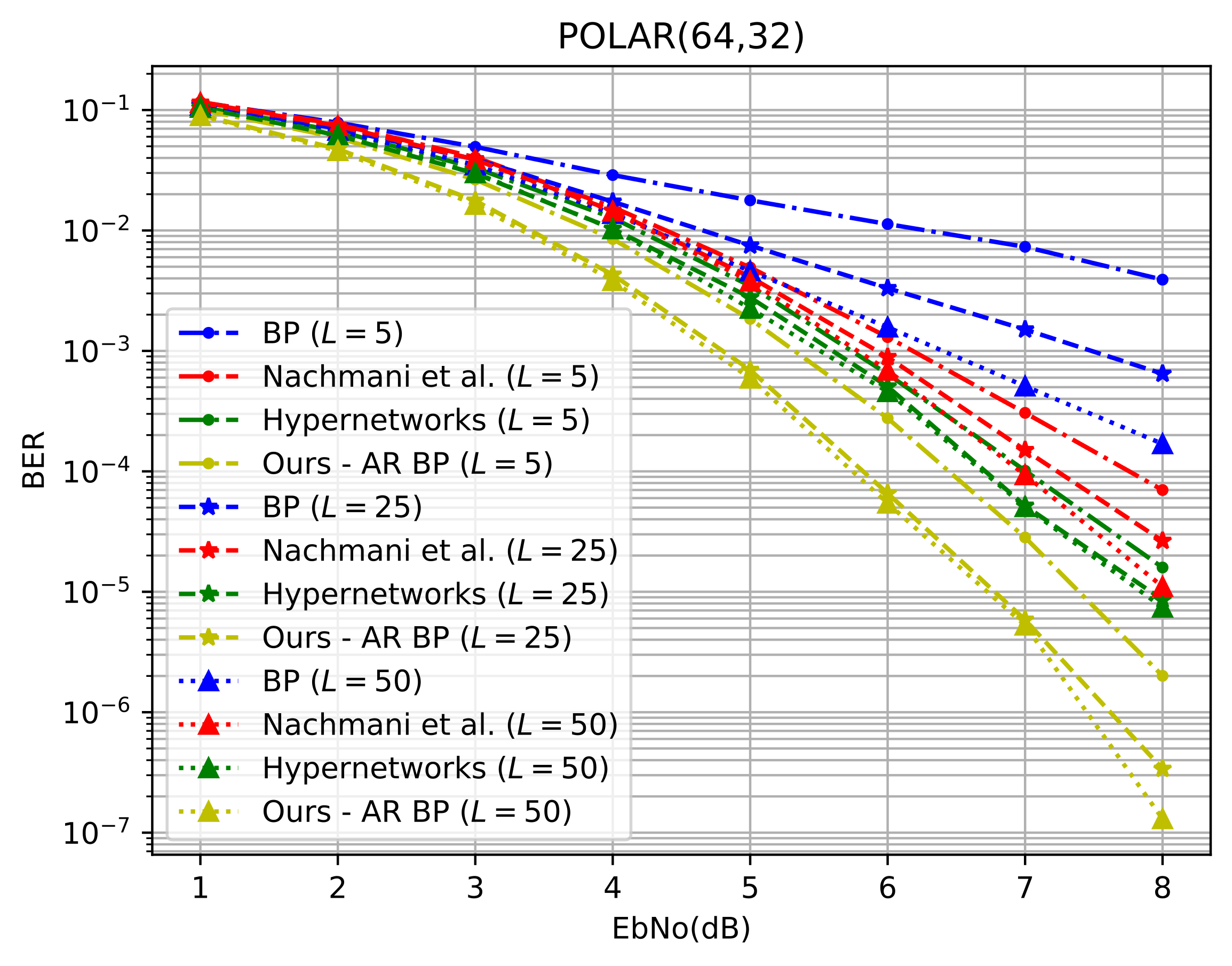}& 
\includegraphics[width=.25\textwidth]{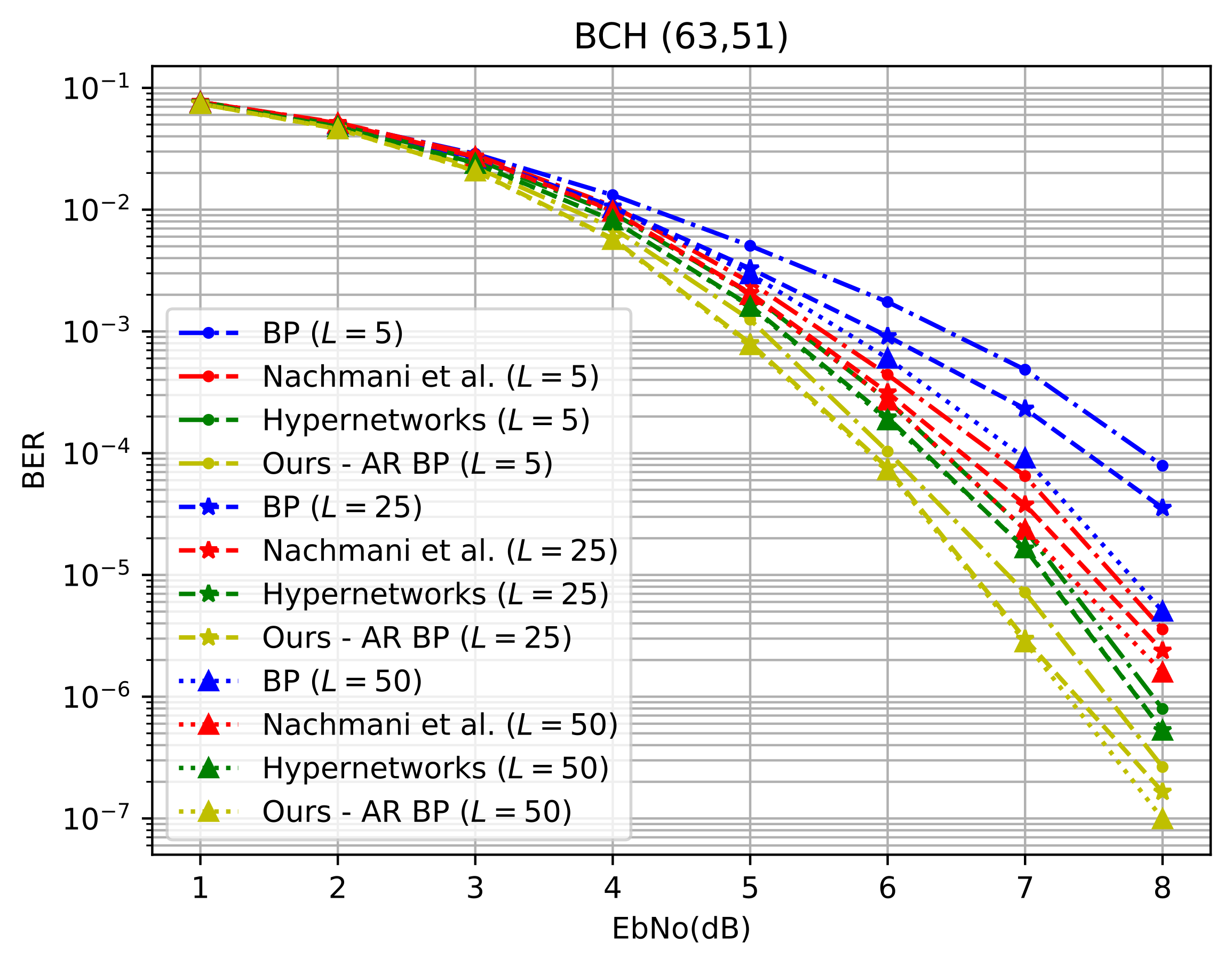} \\
(a) & (b) \\
\end{tabular}
\caption{BER for various values of SNR for various codes. (a) Polar (64,32) and (b) BCH(63,51).}
\label{fig:ber_snr}
\end{figure}

The results are reported in Tab.~\ref{tab:ber_snr}, and are provided for more BER values, for some of the codes in Fig.~\ref{fig:ber_snr}. Fig.~\ref{fig:ber_snr}(a) displays the BER for Polar(64,32) code, where our method achieves an {improvement of $1.2dB$} over~\cite{nachmani2019hyper} for $L=50$. For BCH codes, Fig.~\ref{tab:ber_snr}(b) depict an improvement of $0.5dB$ for BCH(63,51).

Tab.~\ref{tab:ber_snr} presents negative natural logarithm of Bit Error Rate (BER) results for $15$ block codes. Our method obtains after five iterations better results then BP, Learned BP~\cite{nachmani2016learning} and the Hypernetwork BP~\cite{nachmani2019hyper}. The same improved results hold for larger $L$, i.e, convergence of the algorithm.

In order to observe the contribution of the various autoregressive terms and the SNR conditioning, we ran an ablation study. We compare our (i) complete method, (ii) our method without the autoregressive term $a^j$, (iii) our method without the term $e^j$, (iv) our method without $z^j$, (v) our method without $p$ and (vi) our method when trained with noisy variations of zero codeword only. The reduced methods in variants (ii-v) are identical to the complete method, only one term is removed from Eq.~\ref{eq:f_new}.

As can be seen in Tab.~\ref{tab:ablation}, the removal of each of the four novel terms is detrimental. As expected from our analysis, training with the zero codeword is not effective for our method. However, as shown in the same table, for the previous work that maintains the symmetry conditions, training with the zero codeword is slightly better than training with our training set.

\begin{table}[t]
    \centering
    \caption{Ablation analysis. The negative natural logarithm of the Bit Error Rate (BER) at two SNR values for our model five variants of it. Higher is better.}
    \vspace{-0.25cm}
    \label{tab:ablation}
    \setlength\tabcolsep{1.5pt}
    \begin{tabular}{lccccccc}
    \toprule
        & Code & \multicolumn{2}{c}{BCH (31,16)} &  \multicolumn{2}{c}{BCH (63,45)} & \multicolumn{2}{c}{BCH (63,51)}\\
        \cmidrule(lr){3-4}
        \cmidrule(lr){5-6}
        \cmidrule(lr){7-8}
        & Variant/SNR & 7 & 8 & 7 & 8 & 7 & 8\\ 
        \midrule  
        (i) &  Complete method             & 11.94 & 14.50 & 11.48 & 14.08& 12.78 & 16.13 \\
        (ii) & No  $a^j$                   & 11.61 & 13.23 & 10.89 & 13.54 & 11.58 & 14.74 \\
        (iii) & No  $e^j$                  & 11.63 & 13.05 & 10.92 & 13.44 & 11.51 & 14.23 \\
        (iv) & No  $z^j$                   & 11.32 & 13.14 & 10.96 & 13.62 & 11.42 & 14.17 \\
        (v) &  No  $p$                     & 11.57 & 13.18 & 11.01 & 13.23 & 11.98 & 14.55 \\
        (vi) & training with zero codeword & 2.50 & 1.88 & 2.78 & 3.46 & 5.95 &  6.12 \\
        \midrule
        &Hypernet~\cite{nachmani2019hyper} train with zero c.w.  & 10.69 & 13.05 & 11.04 & 14.14 & 11.07 & 13.54\\
        &Hypernet~\cite{nachmani2019hyper} train with random c.w. & 10.47 & 13.01 & 10.13 & 12.89 & 10.23 & 12.44 \\
        \bottomrule 
    \end{tabular}
\end{table}

 \begin{table}[t]
      \centering
            \caption{Runtime per batch in msec}
            \label{tab:runtime}
            \vspace{-2pt}
            \begin{tabular}[b]{p{6.0cm}p{0.5cm}p{0.5cm}} 
            \toprule
            Method          & Train & Test    \\ \midrule
            \cite{nachmani2016learning}    & 16.2  & 5.4   \\ 
            \cite{nachmani2019hyper}      & 88.2  & 40.1  \\ 
            Ours          & 90.9  & 42.6  \\ \bottomrule
            \end{tabular}
\end{table}

\begin{figure}
\centering
\begin{tabular}{c}
\includegraphics[width=0.75\linewidth,height=0.3\textheight,keepaspectratio]{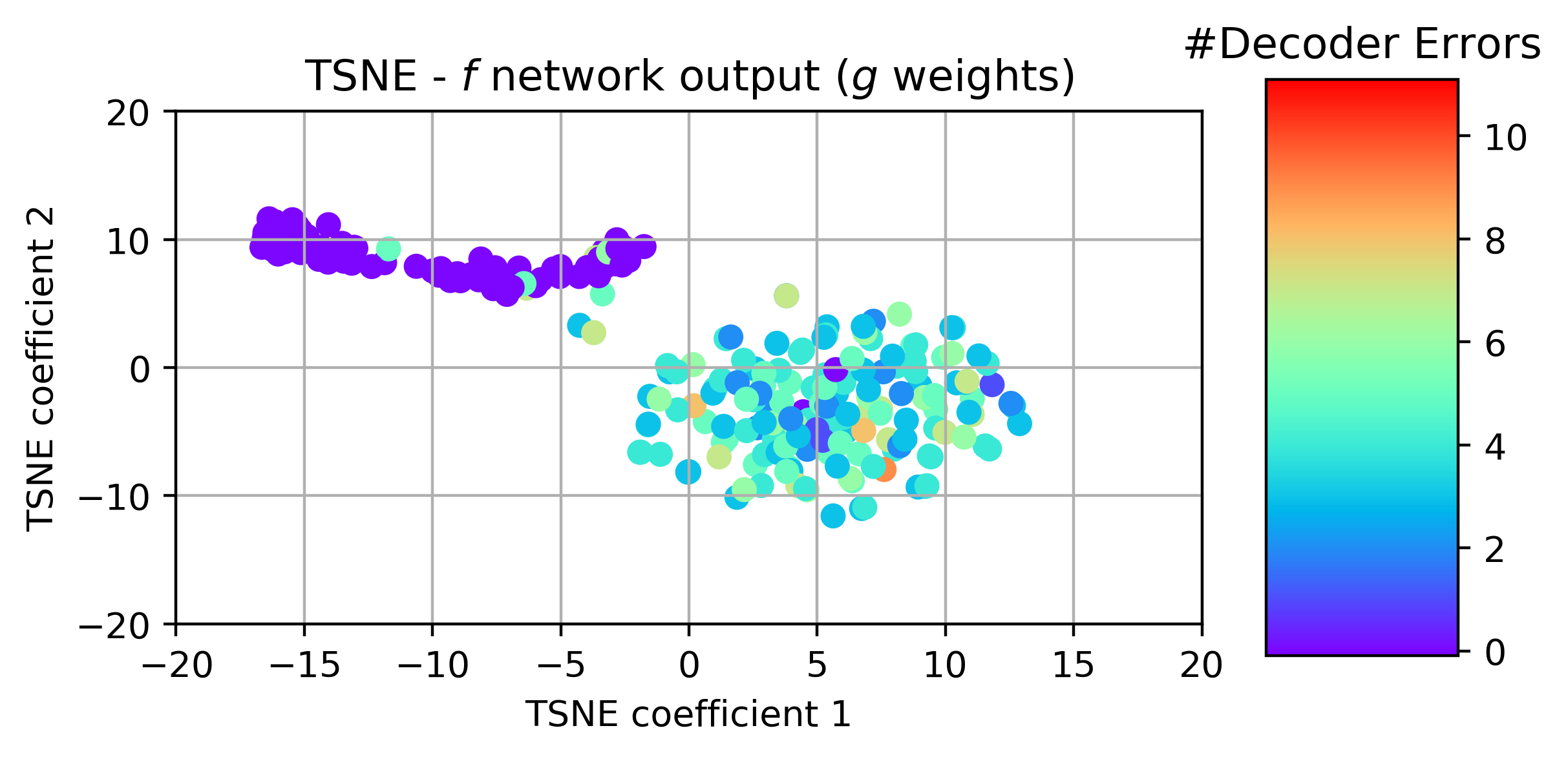}\\ 
(a)\\
\includegraphics[width=0.75\linewidth,height=0.3\textheight,keepaspectratio]{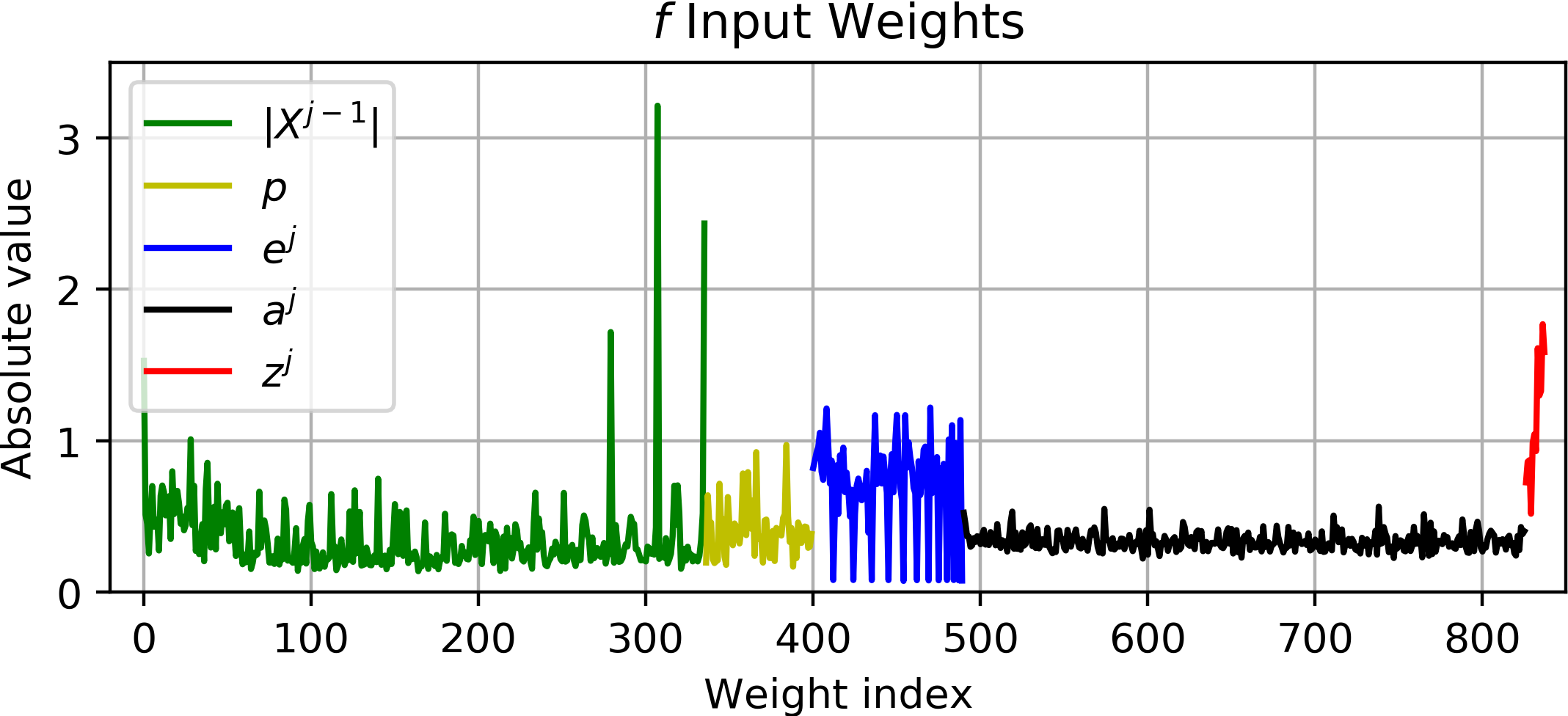} \\
(b) \\
\end{tabular}
\caption{(a) TSNE of $f$ output (b) Abs weight per input}
\label{fig:insight}
\end{figure}

To further understand the behavior of the autoregressive model, we present in Fig.\ref{fig:insight}(a) the TSNE visualization of the output of network $f$ trained for BCH(63,51) code. Each point represents codeword, we color each point with the number of bit errors at the output of the decoder. We can see a clear difference in the primary (dynamic) network $g$ between codewords with significant errors (bottom right) in comparison to words without errors (top left). Fig.\ref{fig:insight}(b) shows the absolute value of the first layer of the network $f$ for the same code. The network emphasizes  $z^{j}$ (shown in red) which is the re-encoding mismatch in Eq.\eqref{eq:z_vec}. The second important part is $e^{j}$ which is the parity check of from matrix $H'$.

\section{Conclusions}
We propose two modifications to the learnable BP methods that currently provide the state of the art results in decoding block codes. First, we embed the estimated SNR, enabling the network to adapt to the level of the incoming noise. Second, we incorporate multiple autoregressive signals that are obtained from the intermediate output of the network.
The usage of autoregressive signals for BP, and graph neural networks in general, can also be beneficial in more domains, such as image completion~\cite{komodakis2007image}, stereo matching~\cite{sun2003stereo}, image restoration \cite{felzenszwalb2006efficient}, codes over graphs \cite{yohananov2019codes} and speech recognition~\cite{hershey2010super}. For example, it would be interesting to see if feeding these networks with computed error and mismatch signals that arise from the intermediate per-iteration solutions can boost performance. 

\newpage
\bibliographystyle{IEEEtran}
\bibliography{literature}

\end{document}